# Preparation of poly(sodium acrylate-co-acrylamide) superabsorbent copolymer via alkaline hydrolysis of acrylamide using microwave irradiation

Hussam-Aldeen Kalaleh, Mohammad Tally and Yomen Atassi*

Laboratory of Materials Sciences, Department of Applied Physics,
Higher Institute for Applied Science and Technology, P.O. Box 31983, Damascus, Syria

* Corresponding Author: yomen.atassi@hiast.edu.sy, Tel. 00 963 11 512 4639

Abstract:
In this paper we present a new one-pot synthesis method of poly(acrylate-co-acrylamide) superabsorbent polymer via partial alkaline hydrolysis of acrylamide using microwave irradiation. This method allows to hydrolysis, polymerization and gelation to take place in one pot during a very short reaction time (90 s), and with no need to operate under inert atmosphere. The degree of hydrolysis of the gel was determined by a back titration method.

The gel is compact and has a water absorbency of 1031 g/g while the corresponding copolymer prepared from polymerization of sodium acrylate and acrylamide, using microwave irradiation and under the same experimental conditions, has a water absorbency of only 658g/g. This difference in water absorbency is discussed. FTIR spectroscopy was used to verify the hydrolysis and the formation of sodium acrylate. Scanning electron microscopy (SEM) showed that the synthesized hydrogel has a macroporous structure. The influence of the environmental parameters on water absorbency such as the pH and the ionic force was also investigated.

1. **Introduction**:

Superabsorbent polymers (SAPs) are three-dimensional cross-linked polymeric structures that are capable of absorbing large quantities of water without disintegrating (Pourjavadi et al. 2009), (Shi et al. 2011), (Pourjavadi et al. 2010). They are widely used in various applications such as hygienic, foods, cosmetics, and agriculture (Brandon & Harland. 1990), (Riccardo. 1994). Therefore the synthesis and the investigation of new SAPs with high absorbency, better mechanical strength and fast initial absorption rate, have been the goal of several research groups in the past decades (Zhao et al. 2005), (Yin et al. 2007). Acrylic-based SAPs have been widely studied because of their good swelling behaviors and chemical stability (Hoffman. 2002).

On the other hand, using microwave irradiation in organic synthesis has received increasing interest due to the specificity of microwave heating in terms of reactivity associated with control of very fast heating rate (Cao et al. 2001), (Stuerga et al. 1993). Microwave energy can be directly and uniformly absorbed throughout the entire volume of the reactive medium, causing it to heat up evenly and rapidly (Danks. 1999). The microwave irradiation has been successfully used in the graft polymerization of chitosan with polyacrylic acid (Huacai et al. 2006), the cellulose on polyacrylamide (Pandey et al. 2013) and bentonite on poly(sodium acrylate-co-acrylamide) (Kalaleh et al. 2013).

The preparation of poly(sodium acrylate-co-acrylamide) with conventional heating methods is very common. It can be prepared from its monomers (Okay et al. 2000), or via alkaline hydrolysis of polyacrylonitrile, polyacrylamide or poly (acrylamide-co-acrylonitrile) (Pourjavadi & Hosseinzadeh. 2010), (Zhao et al. 2010), (Soleimani et al. 2013), (Sadeghi & Hosseinzadeh. 2008).

In this paper we aim to synthesize a high absorbency gel of poly(sodium acrylate-co-acrylamide) in one reaction pot starting from the acrylamide monomer under partial alkaline hydrolysis and using microwave irradiation. The polymerization, the alkaline hydrolysis of amide groups and the cross-linking will take place in one stage and during a very short period of time not exceeding 90 sec. As the reaction time is very short, there's no need to operate under inert atmosphere.

## 2. Experimental
### 2.1. Materials

Acrylamide (AM) for synthesis (Merck) was used as purchased. Potassium peroxodisulfate $K_2S_2O_8$ (KPS) GR for analysis as an initiator and N,N-methylene bisacrylamide (MBA) for electrophoresis as a crosslinker were also obtained from Merck. Sodium hydroxide NaOH microgranular pure (POCH) was used as a hydrolysis agent. Solvents: methanol and ethanol (GR for analysis) were obtained from Merck. Saline sodium chloride (NaCl) (Merck), calcium chloride ($CaCl_2$) (Panreac) and aluminium chloride ($AlCl_3$) (Merck) were prepared with distilled water.

### 2.2. Instrumental Analysis

The IR spectra in the 400-4000 $cm^{-1}$ range were recorded at room temperature on the infrared spectrophotometer (Bruker, Vector 22). For recording IR spectra, powders were mixed with KBr in the ratio 1:250 by weight to ensure uniform dispersion in the KBr pellet. The mixed powders were then pressed in a cylindrical die to obtain clean discs of approximately 1 mm thickness.

Morphology of the dried gels was studied by scanning electron microscope (Vega Tescan SEM). Dried superabsorbent powder was coated with a thin layer of graphite and imaged in a SEM instrument.

Water absorbency of the gel is measured by the free swelling method and is calculated in grams of water per one gram of the dried product (Mohammad et al. 2008). Thus, an accurately weighed quantity of the polymer under investigation (0.1 g) is immersed in 500mL of distilled water at room temperature for at least 3 hours. Then the swollen sample is filtered through weighed 100-mesh (150 μm) sieve until water ceased to drop. The weight of the gel containing absorbed water is measured after draining for 1 hour, and the water absorbency is calculated according to the following equation:

$$Q = (m_2-m_1)/ m_1$$

Where $m_1$ and $m_2$ are respectively the masses (g) of the dry sample and the swollen sample.

### 2.3. Copolymer preparation from acrylamide in alkaline medium (POLY1):

In this section, we aim to prepare poly(sodium acrylate-co-acrylamide), with different molar ratios of acrylate/acrylamide groups, starting from acrylamide in an alkaline medium. To this end, adequate amounts of sodium hydroxide (ranging from 0.02 to 0.08 mol) are added to a solution of acrylamide (0.169 mol in ca. 24mL of distilled water) under vigorous stirring. Then 0.08g of KPS, as an initiator, (in ca. 7 mL of distilled water) and 0.08g of MBA, as a crosslinker, (in ca. 5 mL of distilled water) are added to the above solution. The total mass of the reactive mixture is brought to 100g by adding distilled water. The mixture is stirred for 15 min. Then it is treated in a microwave oven at the power of 950W. Temperature and viscosity of the reactive mixture increase fast. The gelation point is reached after 90s and a recognizable odor of ammonia has accompanied the gel formation. The product (as an elastic gel) is cut to small pieces. Then it is washed several times with methanol to dissolve non reacting reagents, and last washed by ethanol, and dried for several hours at $60^0C$ until it becomes solid and brittle. At this point the solid is ground and dried in the furnace at $60^0C$ for 24 hours.

### 2.4. Determination of the degree of hydrolysis

The hydrolysis degree of each sample is determined by acid-base titration of carboxylate groups produced by alkaline hydrolysis of acrylamide, after adding a saturated solution of NaCl to a precisely pre-weighed mass of the copolymer.

Hydrochloric acid (0.1M) solution is dripped over the sample solution until the pH is reached 3. This solution is then back titrated with aqueous solution of sodium hydroxide (0.1M) until the pH was raised to 7. The number of equivalents of sodium hydroxide consumed is equal to the number of acrylate content of each sample (Rabiee et al. 2005).

### 2.5. Copolymer preparation from sodium acrylate and acrylamide (POLY2):

For a comparative purpose, we aim in this section to synthesize poly(sodium acrylate-co-acrylamide) with the percentage of 18.83% of acrylate groups in the copolymer and starting from its constituting monomers, i.e sodium accrylate and acrylamide. To this end, an adequate amount of acrylic acid is completely neutralized with NaOH solution (2M) in an ice bath to avoid polymerization. This solution is added to acrylamide solution. The total number of moles of monomers is 0.169 moles. Then, the procedure is continued as mentioned above in 2.3. for the synthesis of the copolymer.

## 3. Results and disscussion:
### 3.1. Water absorbency:

Table 1 presents water absorbencies of different samples of POLY1 prepared with different hydrolysis degrees.

| Exp. n$^o$ | number of moles of NaOH (mol) | Water absorbency (g/g) | Gel consistency |
|---|---|---|---|
| 1 | 0 | 34 | compact |
| 2 | 0.02 | 706 | compact |
| 3 | 0.03 | 728 | compact |
| **4** | **0.04** | **1031** | **compact** |
| 5 | 0.05 | 1134 | Slightly compact |
| 6 | 0.06 | 1162 | loose |
| 7 | 0.08 | - | no-gel |

Table 1: Water absorbencies of POLY1 samples with different hydrolysis degrees.

From table 1, we can conclude that the alkaline hydrolysis plays a positive role in increasing water absorbency of the gel, by rendering it hydrophilic and by increasing the inter-chains repulsion due to the electrostatic repulsion of acrylate groups. But increasing NaOH above a certain threshold affects the gel consistency, as the secondary amide group of the crosslinker may undergo alkaline hydrolysis.

In the course of the hydrolysis, the formation of ammonia gas is the driving force that displaces the equilibrium of the alkaline hydrolysis of acrylamide towards the formation of acrylates, and it favors the hydrolysis of the primary amide groups of the acrylamide to the hydrolysis of secondary amide groups of the crosslinker MBA.

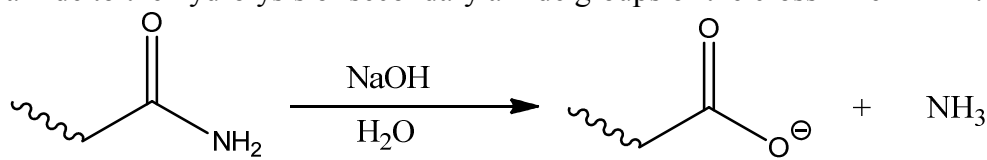

The best water absorbency of POLY1 copolymers is 1031 g/g with a good mechanical consistency of the formed gel. The degree of hydrolysis of this gel, determined according to 2.4., is 18.83%. However, water absorbency of the copolymer POLY2, with the percentage of 18.83% of acrylate groups in the copolymer, prepared from its constituting monomers is 658 g/g.

This difference in water absorpency between the two copolymers POLY1 and POLY2 can be attributed to two factors:
a) The first one is the difference in the placement of the co-monomers along the polymeric network in these two copolymers. In fact, during the hydrolysis of acrylamide, carboxylate groups form randomly along the polymeric chains. So POLY1 is expected to be statistical. On the other hand, the chains of the other copolymer POLY2 resulting from the reaction of the two co-monomers have tendency to blockiness according to the difference of reactivity ratios of those two comonomers (Haque et al. 2010), (Riahinezhad et al. 2013). In the case of the statistical copolymer the electrostatic repulsion of the polymeric chains is more efficient and provides better swelling properties.
b) The second factor is might be linked to the formation of $NH_3$ gas during the alkaline hydrolysis and might be responsible of more porous structure of the gel (Kabiri et al. 2003). In fact, during the reaction in the microwave oven, the reactive mixture becomes jelly or pasty, which prevents the removal of the evolved $NH_3$ and water vapors from the pasty medium (see 2.3.). Therefore, the removed vapors create porosity in the gel. The porosity favors better water diffusion through the hydrogel network.

We have to notice that water absorbencies of the copolymers POLY1 prepared by microwave heating are higher than the water absorbency of the same copolymer when prepared by conventional heating methods (Pourjavadi & Hosseinzadeh. 2010). We assume that microwave heating which induces shorter gelation time (not exceeding 90 s), leads up to more porogen bubbles trapped in the viscous reaction mixture, and results in products with higher porosity (Kabiri et al. 2003).

### 3.2. FTIR spectra:
Figure 1 shows the FTIR spectrum of polyacrylamide gel synthesized without alkaline hydrolysis (A), and the spectrum of the copolymer POLY1 prepared from the alkaline hydrolysis of the acrylamide (B).

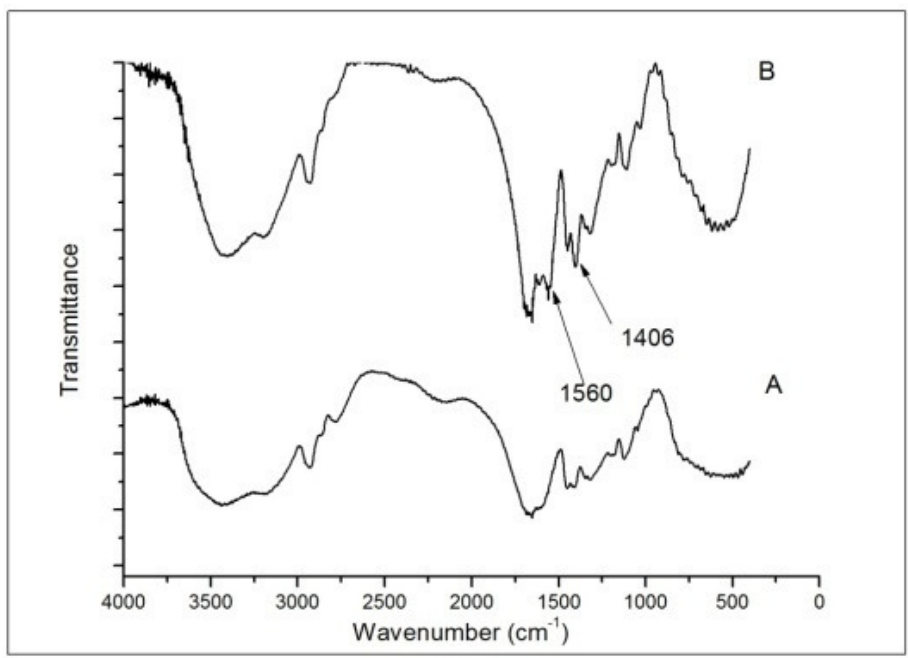

Figure 1 : FTIR spectra of (A) polyacrylamide gel without alkaline hydrolysis and (B) poly (acrylamide–co-sodium acrylate) gel prepared from alkaline hydrolysis of acrylamide.

Both spectra in Figure 1 exhibit the characteristic absorption bands at 1654 cm$^{-1}$ and 3450 cm$^{-1}$, corresponding to the C=O stretching vibration of the amide groups and amide N-H stretching respectively. By comparing the two spectra, it can be seen clearly the appearance, in spectrum (B), of additional two peaks characteristics of the carboxylate ion due to the partials alkaline hydrolysis of the amide groups. The first peak is at about 1406 cm$^{-1}$ and corresponds to the symmetric stretching mode of the carboxylate ion. While the second peak is at 1560 cm$^{-1}$ and corresponds to the asymmetric stretching mode of this ion.

### 3.3. Scanning Electron Microscopy:
One of the most important properties that must be considered when studying hydrogels is its microstructure morphology. Figure 2 shows the scanning electron microscope micrographs of the hydrogel POLY1 with the hydrolysis degree 18.83%.
The micrographs of Figure 2 verify that the synthesized polymer has a porous structure, with the effective pore sizes in the 1-2 μm range. So this copolymer is classifed as a macroporous hydrogel (Ganji et al. 2010). It is supposed that these pores are regions of water permeation and interaction sites between external stimuli and the hydrophilic groups of the gel. These pores were simply produced from $NH_3$ gas formation and water evaporation during hydrogel synthesis.

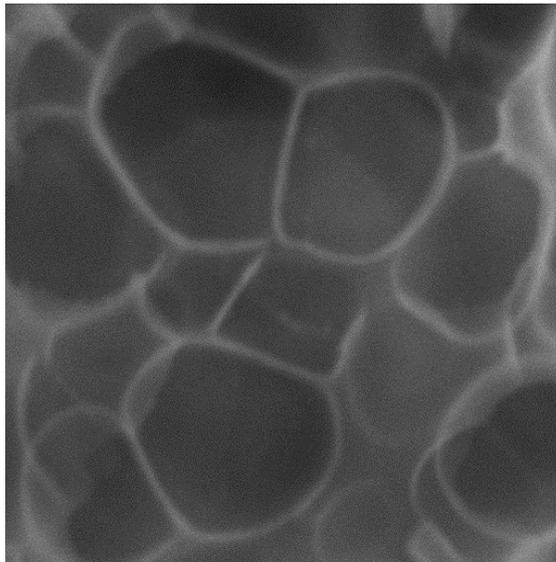
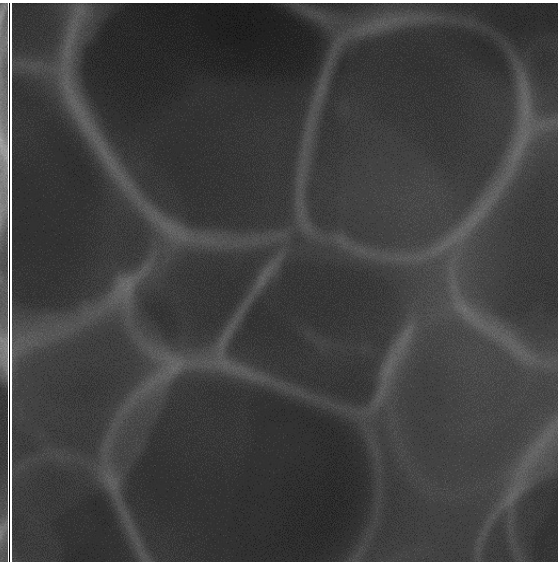
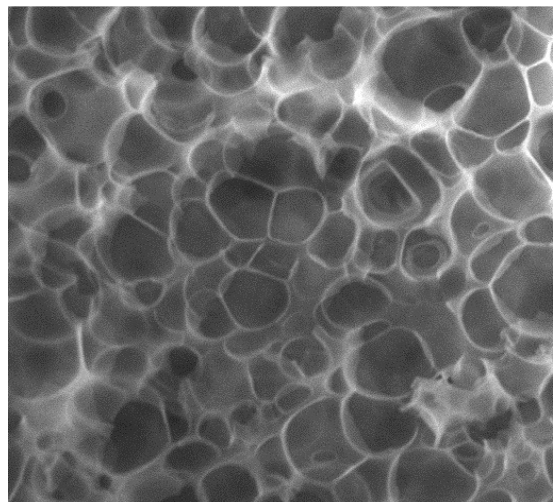
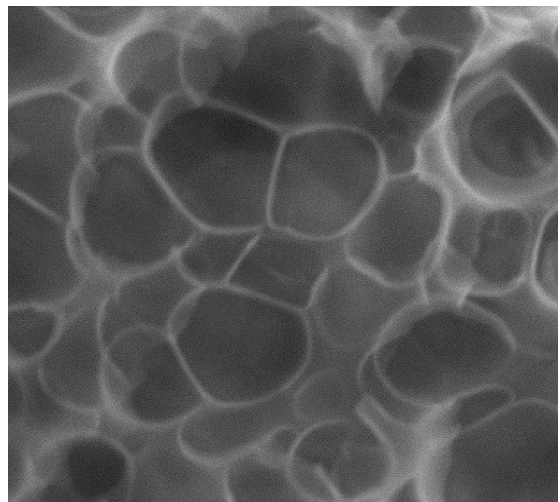
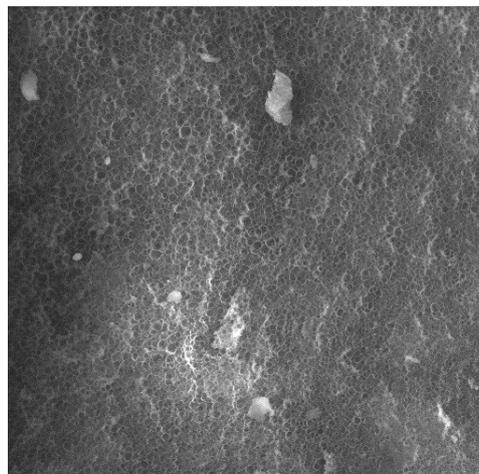

Figure 2: SEM micrographs of the dried poly (acrylamide–co-sodium acrylate) gel prepared from the alkaline hydrolysis of acrylamide (hydrolysis degree 18.83%) with different magnification scales.

### 3.4. Effect of the environmental parameters on water absorbency
#### 3.4.1. Effects of salt solution on water absorbency:

The copolymer POLY1 with the hydrolysis degree 18.83% was tested for the effect of water salinity on its swelling capacity. Different concentration of NaCl, $CaCl_2$, $AlCl_3$ solutions were prepared in order to study the effect of ion charge and ion concentration on water absorption, Figure 3.

Figure 3 shows that water absorption decreases with increasing the ionic strength of the saline solution as cited in Flory equation (Pourjavadi et al. 2005), (Lanthong et al. 2006). The ionic strength of the solution depends on both the concentration and the charge of each individual ion. In fact the presence of ions in the solution decreases the osmotic pressure difference, the driving force for swelling, between the gel and the solution. In addition, multivalent cations ($Ca^{2+}$ and $Al^{3+}$) can neutralize several charges inside the gel by complex formation with carboxamide or carboxylate groups, leading to increased ionic crosslinking degree and consequently loss of swelling.

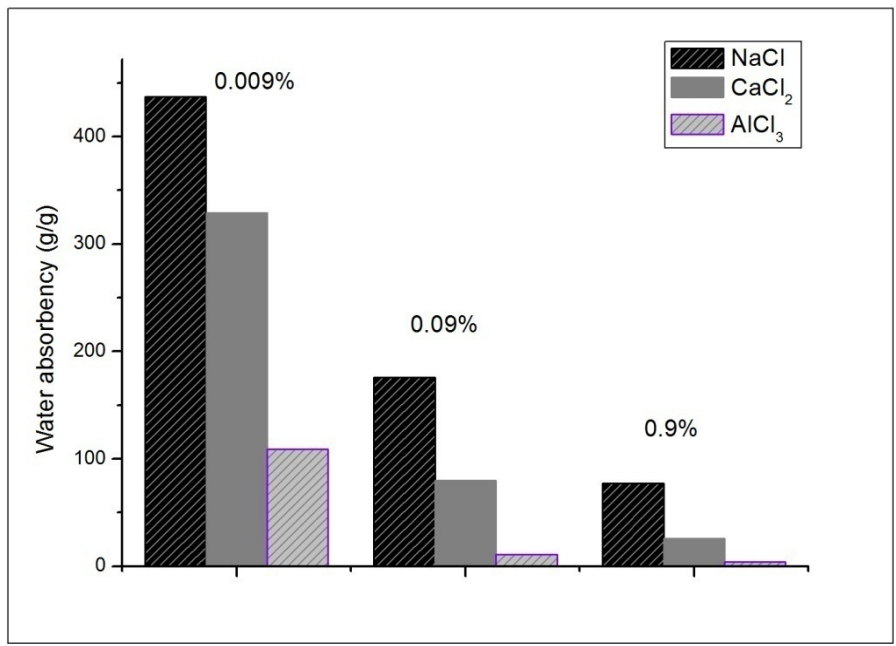

Figure 3: Effect of salt solution on water absorbency of the copolymer POLY1 with the hydrolysis degree 18.83%.

#### 3.4.2. Effects of the environmental pH on water absorbency:

The swelling behavior of the sample POLY1 with the hydrolysis degree 18.83% was studied at various pH values between 2.0 and 12.0, at room temperature, Figure 4. To clearly observe the net effect of pH, buffer solutions containing lots of ionic species, were not used as swelling media, as the swelling is strongly decreased by ionic strength. Therefore, a stock of concentrated solution HCl and NaOH were diluted with distilled water to reach the desired acidic or basic pH.

Studies have indicated that water absorption of hydrogels are sensitive to environmental pH (Sunny et al. 2009). Figure 4 shows that at low pH, the swelling capacity decreases as the sodium carboxylate groups on the polymer network are protonated. The polymer shrinks and becomes hydrophobic. This in turn decreases the degree of ionization and hence decreases the swelling ratio.

At high pH, the swelling capacity also decreases by "charge screening effect" of excess $Na^+$ in the swelling media, which shields the carboxylate anions and prevents effective anion-anion repulsion (Sunny et al. 2009).

In the interval pH=4 to pH=8, some of carboxylic acid groups are ionized and the electrostatic repulsion between $COO^-$ groups causes an enhancement of the swelling capacity.

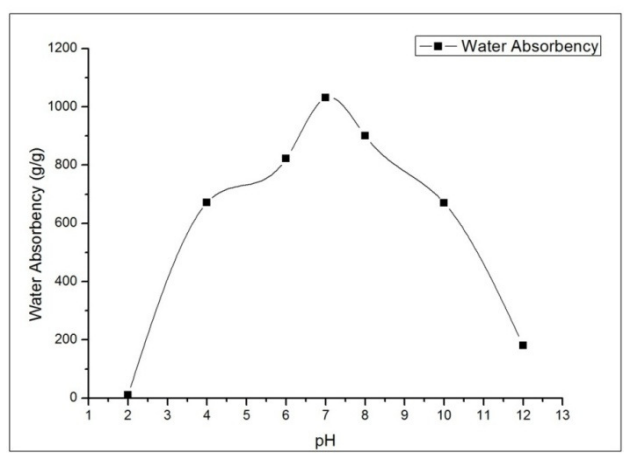

Figure 4: Effect of environmental pH on water absorbency of the copolymer POLY1 with the hydrolysis degree 18.83%.

4. **Conclusion**

Two types of poly(sodium acrylate-co-acrylamide) have been synthesized by a simple and fast method, using a domestic microwave oven. One of them was prepared from partially hydrolyzed acrylamide and the other one from its corresponding co-monomers. The difference in water absorbencies between the copolymers was linked on one hand, to the statistical nature of the first copolymer compared with the other one which was a block copolymer. On the other hand, the more porous structure of the statistical copolymer, due to gassing of ammonia, favors more water absorption. The relatively high values of water absorbency of these copolymers compared to the absorbency of similar copolymers prepared via classical heating are related to the very short gelation time induced by microwave heating that leads up to more porogen bubbles trapped in the viscous reaction mixture, and results in products with higher porosity. SEM micrographs demonstrate the macroporous nature of the gel.

**Figure and Table Legends:**
Figure 1 : FTIR spectra of (A) polyacrylamide gel without alkaline hydrolysis and (B) poly (acrylamide–co-sodium acrylate) gel prepared from alkaline hydrolysis of acrylamide.
Figure 2: SEM micrographs of the dried poly (acrylamide–co-sodium acrylate) gel prepared from the alkaline hydrolysis of acrylamide (hydrolysis degree 18.83%) with different magnification scales.
Figure 3: Effect of salt solution on water absorbency of the copolymer POLY1 with the hydrolysis degree 18.83%.
Figure 4: Effect of environmental pH on water absorbency of the copolymer POLY1 with the hydrolysis degree 18.83%.

Table 1: Water absorbencies of POLY1 samples with different hydrolysis degrees.